\documentclass[10pt,conference]{IEEEtran}

\ifCLASSINFOpdf
   \usepackage[pdftex]{graphicx}
\else
   \usepackage[dvips]{graphicx}
\fi

\ifCLASSOPTIONcomsoc
  \usepackage[caption=false,font=normalsize,labelfont=sf,textfont=sf]{subfig}
\else
  \usepackage[caption=false,font=footnotesize]{subfig}
\fi

\usepackage{footnote}
\usepackage{nopageno}
\usepackage{float}
\usepackage{url}
\usepackage{amssymb,amsfonts}
\usepackage{amsmath}

\usepackage{graphicx}
\usepackage{mathptmx}  
\usepackage{epstopdf}
\usepackage{textcomp}
\usepackage{xcolor}
\usepackage{xspace}
\usepackage[absolute,showboxes]{textpos}

\usepackage{multirow}
\usepackage{multicol}
\usepackage{array} 
\usepackage{cite}
\usepackage{algorithm,algorithmic}
\usepackage{amsmath}
\usepackage[cmintegrals]{newtxmath}

\usepackage[utf8]{inputenc}
\usepackage{blindtext}
\usepackage{csquotes}

\begin{document}
\pagenumbering{gobble}

\title{DeepStage: Learning Autonomous Defense Policies Against Multi-Stage APT Campaigns}

\author{

Trung V. Phan\IEEEauthorrefmark{2}, Tri Gia Nguyen\IEEEauthorrefmark{4} and Thomas Bauschert\IEEEauthorrefmark{2} \\
\IEEEauthorblockA{\IEEEauthorrefmark{2}Chair of Communication Networks, Technische Universit{\"a}t Chemnitz,  09126 Chemnitz, Germany}
\IEEEauthorblockA{\IEEEauthorrefmark{4}Department of Information Assurance, FPT University, Da Nang 50509, Vietnam}
Email: trung.phan-van@etit.tu-chemnitz.de, tri@ieee.org, thomas.bauschert@etit.tu-chemnitz.de


}

\maketitle

\begin{abstract}
This paper presents \textit{DeepStage}, a deep reinforcement learning (DRL) framework for adaptive and stage-aware defense against Advanced Persistent Threats (APTs). The enterprise environment is formulated as a partially observable Markov decision process (POMDP), in which host provenance and network telemetry are fused into unified provenance graphs. Building on our prior work~\cite{StageFinder}, \textit{DeepStage} employs a graph neural network encoder and an LSTM-based stage estimator to infer probabilistic attacker stages aligned with the MITRE ATT\&CK framework. The resulting stage beliefs, together with graph embeddings, are used to guide a hierarchical Proximal Policy Optimization (PPO) agent that selects defense actions across monitoring, access control, containment, and remediation. Experiments in a realistic enterprise testbed with CALDERA-driven APT playbooks show that \textit{DeepStage} achieves an average F1-score of 0.887 and a mitigation success rate of 84.7\%, outperforming a risk-aware DRL baseline by 21.8\% in F1-score and 16.2\% in mitigation success. The results demonstrate effective stage-aware and cost-efficient autonomous cyber defense.

\end{abstract}

\IEEEpeerreviewmaketitle

\pagestyle{headings}
\setcounter{page}{1}
\pagenumbering{arabic}

\section{Introduction}\label{Introduction}
In recent years, Advanced Persistent Threats (APTs)~\cite{APTDetectionSurvey1} have emerged as a major security concern for enterprise, governmental, and critical infrastructure networks. Unlike opportunistic or commodity malware, APTs are characterized by stealthy operations, prolonged dwell times, and multi-stage attack progressions aimed at achieving long-term objectives such as data exfiltration, cyber espionage, or operational sabotage. As reflected in the MITRE ATT\&CK Enterprise matrix~\cite{mitre_attack}, an APT campaign typically evolves through multiple stages, beginning with reconnaissance and initial compromise, followed by privilege escalation, lateral movement, and ultimately data exfiltration or system disruption. Each stage often generates only subtle indicators that are interleaved with benign system activity, making reliable detection particularly challenging~\cite{APTDetectionSurvey2}. Traditional signature-based intrusion detection and prevention systems (IDS/IPS) remain effective against known threats but struggle to identify novel or evolving Tactics, Techniques, and Procedures (TTPs)~\cite{DeepLearningforAPTDetection}. Anomaly-based detection approaches provide broader coverage; however, they often suffer from high false-positive rates and limited ability to contextualize sequences of activities across time and hosts~\cite{IDSleveragingHostData}.

To address these challenges, our prior work introduced \textit{StageFinder} \cite{StageFinder}, a framework designed to estimate the current attack stage during an APT campaign. StageFinder continuously collects host-level system logs and transforms them into provenance graphs that capture causal and temporal dependencies among system entities such as processes, files, users, and sockets. To incorporate broader situational context, an early fusion mechanism integrates network-layer alerts—such as those generated by IDS or firewall systems—directly into the provenance graph. Each alert is modeled as a first-class node connected to relevant host entities, preserving semantic relationships between network anomalies and local activities. The resulting fused provenance graph captures both intra-host and inter-network dependencies within a unified causal representation. A graph neural network encoder is then applied to extract low-dimensional embeddings that represent structural and contextual patterns within the graph. Finally, these embeddings are processed by a long short-term memory model to capture temporal dynamics and infer the attacker’s probabilistic stage in the kill chain.

As a continuation of our efforts to defend against APT attacks, this work proposes \textit{DeepStage}, a unified and stage-aware APT defense framework that integrates provenance graph embeddings with deep reinforcement learning (DRL). DeepStage builds upon our prior framework, \textit{StageFinder} \cite{StageFinder}, by inheriting its network and system data processing pipeline as well as its output of probabilistic attack-stage estimates. Specifically, the inferred attack-stage probabilities, together with provenance graph embeddings, are used to condition a hierarchical DRL agent. The agent operates under a partially observable Markov decision process (POMDP) formulation, enabling adaptive and stage-specific defense decisions in the presence of uncertainty and incomplete system visibility. Furthermore, the DRL agent has direct access to both system-level provenance signals and network-level alerts, allowing it to make timely and context-aware defensive responses. Experiments conducted in a realistic enterprise testbed using CALDERA-driven APT playbooks~\cite{mitre_caldera} demonstrate that DeepStage achieves a stage-weighted F1-score of 0.887 and a mitigation success rate of 84.7\%, outperforming a risk-aware DRL baseline in overall APT defense effectiveness.

\section{Related Work}

\subsection{DRL-based Autonomous Cyber Defense}

Deep Reinforcement Learning (DRL) has recently been explored for automated cyber-defense, where an agent learns response policies through interaction with a simulated security environment~\cite{DRLforCyberSecurity}. These systems typically model network defense as a sequential decision problem and train agents to select mitigation actions that balance detection effectiveness, response latency, and operational cost. 

Several DRL-based network intrusion response systems (NIRS) have been proposed to automate incident response~\cite{DRL-basedNIRS}. In these frameworks, the environment is represented by network alerts or host status indicators, and the DRL agent learns to deploy countermeasures such as traffic filtering or service isolation. However, most existing NIRS frameworks are designed for single-phase attacks (e.g., denial-of-service or network scanning) and operate primarily at the network perimeter. Consequently, they lack visibility into host-level behavior and cannot capture the sequential progression of Advanced Persistent Threats (APTs). As a result, their learned policies are largely reactive and respond to individual alerts rather than anticipating the evolution of multi-stage attacks.

\subsection{Risk-Aware DRL and Attack-Graph-Based Defense}

To improve situational awareness, several studies integrate attack-graph analysis with reinforcement learning to model the security state of enterprise networks~\cite{AutomatedAPTDefense}. In these approaches, nodes represent vulnerabilities or privilege states, and edges encode potential attack paths. The DRL agent selects defensive actions that minimize the expected cumulative attack risk, often formulated as $\Delta(\text{total risk}) - \lambda C(a)$, where $C(a)$ denotes the operational cost of action $a$. 

While attack-graph-based methods provide structured reasoning about attacker movement, they rely on static, precomputed graphs generated by tools such as MulVAL. These representations do not capture temporal dependencies among system events and cannot reflect the dynamic and adaptive behaviors characteristic of real-world APT campaigns. Recent proactive DRL approaches attempt to predict attacker movement along possible attack paths~\cite{DRLbasedAttackPathPrediction}. However, these methods typically rely on coarse abstractions of network topology or host connectivity and therefore lack the fine-grained behavioral context required to distinguish concurrent attack stages.

\subsection{Hierarchical and Constrained Reinforcement Learning}

Hierarchical DRL architectures have been proposed to address the complexity of large action spaces in cyber defense. DeepShield~\cite{DeepShield}, for example, models reconnaissance-phase defense as a hierarchical control problem. A meta-agent detects suspicious scanning behavior and selects among several mitigation strategies executed by a lower-level agent, such as IP shuffling, software diversity, or component redundancy. Although hierarchical control improves scalability, DeepShield focuses primarily on early reconnaissance attacks and does not extend to later APT stages such as privilege escalation, lateral movement, or data exfiltration.

Other approaches explore offline or constrained reinforcement learning for cyber defense~\cite{OfflineRLforAutonomousDefense}. These methods aim to learn safe policies from historical security logs while enforcing operational constraints. However, they generally do not incorporate explicit modeling of attack stages, limiting both interpretability and the ability to adapt defensive strategies across different phases of an attack campaign.

\subsection{Limitations of Existing DRL-based Defense Systems}

A key limitation of most DRL-based defense frameworks~\cite{DRL-basedNIRS,AutomatedAPTDefense,DRLbasedAttackPathPrediction,OfflineRLforAutonomousDefense,DeepShield} is their reliance on flat feature-based state representations. Typically, the system state is represented using aggregated indicators such as alert counts, vulnerability scores, or host states. These representations ignore the causal relationships among processes, files, users, and network connections. As a result, the learned policies cannot reason about how malicious activities propagate through system dependencies or evolve across multiple hosts during an APT campaign.

Provenance graphs provide a richer representation for modeling system behavior. By capturing causal dependencies and temporal relationships among system entities, provenance graphs enable fine-grained reasoning about attack progression and system compromise. Embedding such graphs into compact vector representations allows learning-based models to incorporate both structural and temporal context.

\subsection{Summary and Research Gap}

Although prior DRL-based cyber-defense systems demonstrate the potential for automated mitigation, they remain limited in handling multi-stage APT campaigns. Attack-graph-based methods rely on static abstractions, hierarchical DRL frameworks focus on early-stage attacks, and feature-based models fail to capture causal dependencies in system activity. 

To address these limitations, this paper proposes \textit{DeepStage}, a stage-aware defense framework that integrates fused host–network provenance graph embeddings with an LSTM-based stage estimator. Combined with a hierarchical Proximal Policy Optimization (PPO) agent, DeepStage enables adaptive, stage-aware defense policies that respond to the evolving tactics of advanced persistent threats.

\begin{figure*}
    \centering
    \includegraphics[width=1.0\linewidth]{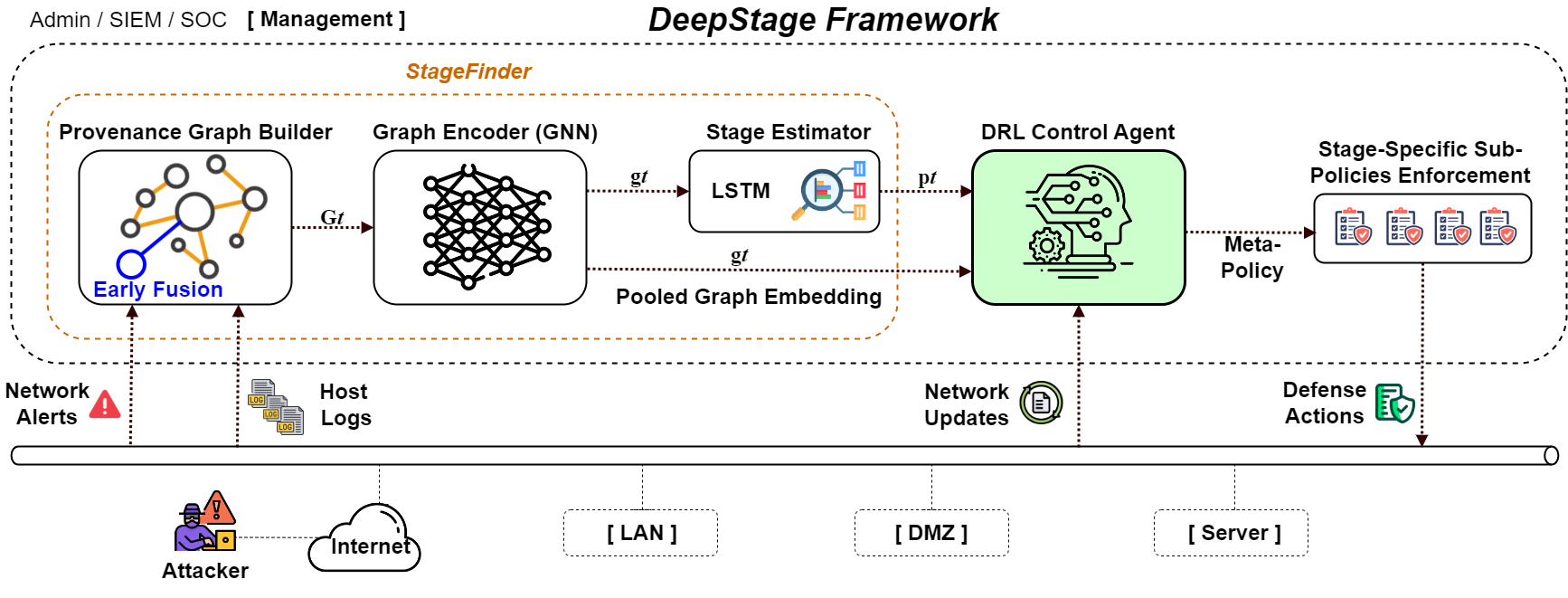}
    \caption{Data and control flow of the proposed DeepStage framework.}
    \label{fig:framework_flow}
\end{figure*}

\section{Design of The DeepStage Framework}\label{ProposedFramework}
\subsection{Network Environment}
We consider a representative enterprise network environment for data collection and analysis. The infrastructure is logically segmented into four zones—local area network (LAN), demilitarized zone (DMZ), server zone, and management zone—as illustrated in Fig.~1. The LAN hosts employee workstations and internal services, which often serve as initial entry points for attacks. The DMZ contains externally accessible services (e.g., web, email, and VPN) and acts as a buffer between external and internal networks. The server zone stores critical assets such as database and authentication servers, frequently targeted in later stages of APT campaigns. The management zone hosts centralized monitoring and orchestration components, where the \textit{DeepStage} framework is deployed to collect telemetry from all zones while remaining isolated from external access. Network traffic is regulated by perimeter firewalls and analyzed by IDS/IPS sensors (e.g., Zeek), enabling the collection of both host- and network-level telemetry for subsequent data fusion.

\subsection{Operation of the DeepStage Framework}\label{sec:DeepStageOperation}

The proposed \textit{DeepStage} framework operates as a closed-loop APT defense architecture that continuously collects system telemetry, models system behavior, and enables adaptive defensive actions. Figure~\ref{fig:framework_flow} illustrates the data and control flow across the system processes.

\subsubsection{Data Acquisition}
Two primary data streams are collected across the enterprise network: (\textit{i}) \textit{host-level logs} from endpoint monitoring tools such as \emph{Auditd} (Linux) and \emph{Sysmon} (Windows), capturing events including process creation and file access; and (\textit{ii}) \textit{network alerts} generated by security sensors such as Snort or Zeek~\cite{zeek_ids}. All events are securely transmitted to the Management Zone for analysis.

\subsubsection{Provenance Graph Construction}
The collected logs and alerts are parsed and correlated to construct a \textit{fused provenance graph}. Nodes represent system entities (e.g., processes, files, sockets, or alerts), while edges encode causal or data-flow relationships. Network alerts are incorporated through an early-fusion mechanism, linking network events with host activities to form a unified causal representation.

\subsubsection{Graph Embedding and Stage Estimation}
A graph neural network (GNN) encoder transforms the fused graph into a fixed-length embedding $g_t$, representing the system state at time $t$. A sequence of embeddings $\{g_1,\ldots,g_t\}$ is then processed by an LSTM-based stage estimator \cite{StageFinder} that outputs a probability distribution $p_t$ over APT stages defined by the MITRE ATT\&CK framework.

\subsubsection{Hierarchical DRL Defense}
The combined state $s_t=[g_t,p_t]$ is provided to a hierarchical DRL agent. The meta-policy interprets the estimated stage and selects a corresponding sub-policy to execute defensive actions such as host isolation, IP blocking, honeypot deployment, or alert escalation.

\subsubsection{Feedback and Learning}
Following each action, the environment provides a reward reflecting mitigation effectiveness and system stability. This feedback updates the DRL policy, enabling continuous adaptation to evolving APT behaviors.

\section{Stage-Aware DRL-Based APT Defense}\label{DRL-BasedAPTDefense}

\subsection{POMDP-Based System Model}
Advanced Persistent Threats are stealthy and multi-stage attacks whose activities are only partially observable through system logs and network alerts. Since the defender cannot directly observe the true system security state, the defense problem violates the full observability assumption of standard Markov Decision Processes (MDPs). Therefore, we model the defense environment as a \textit{Partially Observable Markov Decision Process} (POMDP)~\cite{POMDP}. 

Formally, the environment is defined as 
$\mathcal{M}=(\mathcal{S},\mathcal{A},\mathcal{O},P,\Omega,R,\gamma)$, 
where $\mathcal{S}$ denotes latent system states, $\mathcal{A}$ the defense action space, and $\mathcal{O}$ the observation space. $P(s'|s,a)$ represents state transitions, $\Omega(o|s,a)$ the observation model, $R(s,a)$ the reward function, and $\gamma$ the discount factor. This formulation enables the agent to reason over uncertain system states using belief representations derived from observable evidence.

\subsection{State Representation}
The hidden system state $s_t$ reflects the underlying security posture of the enterprise network, including host compromise levels and the attacker’s progression stage. Because $s_t$ cannot be directly observed, the agent maintains a belief state derived from observable telemetry.

At each time step, host logs and network alerts are fused into a provenance graph $G_t$. A graph neural network (GNN) encoder produces a structural embedding
$g_t = f_{\text{GNN}}(G_t)$, which summarizes system activity. Meanwhile, the Stage Estimator generates a probabilistic stage vector $p_t = [p_t^{(0)}, p_t^{(1)}, \ldots, p_t^{(K)}]$, where $p_t^{(k)}$ denotes the likelihood of APT stage $k$ and $k=0$ represents benign activity. 

The observation at time $t$ is defined as $o_t=[g_t,p_t,a_{t-1}]$, and a recurrent encoder updates the belief representation $b_t=f_{\text{LSTM}}(b_{t-1},o_t)$. The belief embedding $b_t$ integrates structural system context and stage inference, forming the state representation used by the DRL agent.

\subsection{Defense Action Space}
The action space $\mathcal{A}$ consists of practical countermeasures used in enterprise incident response:
\[
\mathcal{A}=\mathcal{A}_{mon}\cup\mathcal{A}_{acc}\cup\mathcal{A}_{cont}\cup\mathcal{A}_{rem}.
\]

\textit{Monitoring ($\mathcal{A}_{mon}$)}: Passive actions that increase situational awareness without disrupting operations, such as increasing log verbosity, triggering malware scans, or correlating alerts.

\textit{Access Control ($\mathcal{A}_{acc}$)}: Defensive actions that restrict privileges or authentication, including disabling suspicious accounts, revoking session tokens, or enforcing multi-factor authentication.

\textit{Containment ($\mathcal{A}_{cont}$)}: Active interventions that limit attacker movement, such as isolating compromised hosts, blocking malicious IPs, or terminating suspicious processes.

\textit{Remediation ($\mathcal{A}_{rem}$)}: Recovery actions that restore system integrity, including patching vulnerabilities, restoring configurations, and resetting security policies.

This structured action space enables a hierarchical DRL agent to activate stage-specific sub-policies aligned with enterprise defense workflows.

\subsection{Reward Design}
The reward function balances two objectives: maximizing attack mitigation effectiveness and minimizing operational disruption. The reward at time $t$ is defined as
\begin{equation}
R(b_t,a_t)=R_{\text{security}}(t)-\lambda C(a_t),
\end{equation}
where $R_{\text{security}}$ measures the effectiveness of the chosen action in blocking or containing malicious activity, $C(a_t)$ represents the operational cost of executing action $a_t$, and $\lambda$ is a global trade-off coefficient controlling the overall penalty strength.

Because different APT stages pose different levels of risk, we introduce stage-aware weighting:
\begin{equation}
R(b_t,a_t)=\alpha_k R_{\text{security}}(t)-\beta_k\lambda C(a_t),
\end{equation}
where $\alpha_k$ and $\beta_k$ adjust the relative importance of security and operational cost at stage $k$. Typically, $\alpha_k$ increases for later stages (e.g., exfiltration), allowing more aggressive responses, while $\beta_k$ decreases to reduce the penalty for disruptive containment actions.

\subsection{Optimization Objective}
The goal of the defense agent is to learn a policy that maximizes the expected cumulative discounted reward:
\begin{equation}
J(\theta)=
\mathbb{E}_{\pi_\theta}
\left[
\sum_{t=0}^{T}\gamma^t R(b_t,a_t)
\right],
\end{equation}
where $\pi_\theta(a_t|b_t)$ is the policy parameterized by neural network weights $\theta$.

\subsection{Hierarchical Policy Structure}
To reflect the multi-stage nature of APTs, we employ a hierarchical policy composed of a meta-policy and stage-specific sub-policies:
\begin{equation}
\pi_{\theta}(a_t|b_t)=
\pi_{\theta_{\text{meta}}}(k_t|b_t)\,
\pi_{\theta_k}(a_t|b_t),
\end{equation}
where the meta-policy selects the current stage $k_t$, and the corresponding sub-policy determines the defense action.

\subsection{Policy Learning via PPO}
To optimize the hierarchical policy, we adopt Proximal Policy Optimization (PPO)~\cite{PPO_Algorithm}, a stable policy-gradient algorithm suitable for high-dimensional and partially observable environments. PPO updates the policy parameters using the gradient
\begin{equation}
\nabla_\theta J(\theta)=
\mathbb{E}_{\pi_\theta}
\left[
\nabla_\theta \log \pi_\theta(a_t|b_t)\,A_t
\right],
\end{equation}
where $A_t=Q(b_t,a_t)-V(b_t)$ denotes the advantage function. PPO enables stable policy updates while improving sample efficiency in complex cyber-defense environments.

\section{Performance Evaluation}\label{PerformanceEvaluation}

\begin{table*}
\centering
\caption{Set of practical sub-policies deployable. Cost $C(a)$ denotes the normalized operational impact.}
\label{tab:subpolicies}
\small
\setlength{\tabcolsep}{4pt}
\begin{tabular}{p{1.0cm} p{5.5cm} p{9.0cm} c}
\hline
\textbf{Meta-policy} & \textbf{Sub-policy} & \textbf{Deployment in Testbed} & \textbf{Cost $C(a)$} \\
\hline
$\mathcal{A}_{\text{mon}}$ 
& $a_{0}$: Maintain baseline monitoring & Keep \textit{auditd.conf}, CamFlow, and Zeek agent in baseline mode & 0.01 \\
& $a_{1}$: Increase host logging level & Modify \textit{auditd.conf}, CamFlow, and Zeek agent policies to capture process and socket events & 0.05 \\
& $a_{2}$: Activate Deep Packet Inspection & Temporarily enable Zeek full packet capture on target subnets & 0.10 \\
& $a_{3}$: Memory/process snapshot & Use \textit{volatility} or \textit{psrecord} via management API to collect runtime memory dump & 0.20 \\
& $a_{4}$: Deploy honeypot redirect & Deploy Cowrie SSH honeypot or redirect via \textit{iptables DNAT} & 0.25 \\
& $a_{5}$: Log enrichment with threat intel & Query local MISP feed and annotate alerts with TTP tags in provenance database & 0.10 \\
& $a_{6}$: Trigger system audit scan & Run \textit{osqueryi} snapshot to collect running processes, loaded modules, and sockets & 0.15 \\
& $a_{7}$: Initiate cross-host correlation scan & Run graph-matching between host provenance graphs to detect coordinated anomalies & 0.20 \\
\hline
$\mathcal{A}_{\text{acc}}$ 
& $a_{8}$: Disable risky services & Stop insecure services (\textit{telnetd}, \textit{ftp}) via \textit{systemctl} & 0.20 \\
& $a_{9}$: Rotate user credentials & Force password reset or SSH key rotation through LDAP/AD & 0.30 \\
& $a_{10}$: Enforce 2FA/MFA on privileged accounts & Integrate PAM-based OTP for sudo/SSH login attempts & 0.40 \\
& $a_{11}$: Revoke active sessions & Terminate processes with elevated privileges & 0.25 \\
& $a_{12}$: Privilege escalation blocklist & Apply AppArmor/Seccomp policy limiting setuid/setcap operations & 0.20 \\
& $a_{13}$: Enable strict sudo logging & Configure \textit{/etc/sudoers} for audit logs and restricted group access & 0.15 \\
& $a_{14}$: Lock compromised accounts & Disable login for suspicious users (\textit{usermod -L}) & 0.30 \\
\hline
$\mathcal{A}_{\text{cont}}$ 
& $a_{15}$: Block malicious IP or domain & Apply \textit{iptables} or Zeek blacklist rule; block known C2 endpoints & 0.20 \\
& $a_{16}$: Throttle suspicious network flows & Use \textit{tc qdisc} or SDN API to rate-limit outgoing traffic & 0.25 \\
& $a_{17}$: Micro-segmentation & Apply dynamic bridge rules to limit host-to-host communication & 0.35 \\
& $a_{18}$: Kill malicious process & Send \textit{SIGKILL}/\textit{SIGSTOP} to identified malicious PID & 0.20 \\
& $a_{19}$: Network interface isolation & Temporarily down network interface (\textit{ifdown eth0}) for suspect host & 0.80 \\
& $a_{20}$: Contain file I/O access & Apply read-only permissions to sensitive directories via \textit{chattr +i} & 0.40 \\
& $a_{21}$: Block USB/external device usage & Disable \textit{usb-storage} kernel module & 0.50 \\
\hline
$\mathcal{A}_{\text{rem}}$ 
& $a_{22}$: Emergency patching & Run \textit{apt-get upgrade --only-upgrade} to patch known CVEs & 0.40 \\
& $a_{23}$: Remove persistence artifacts & Delete malicious autoruns, crontab, rc scripts, or binary replacements & 0.30 \\
& $a_{24}$: Rollback system to snapshot & Use OpenStack/QEMU snapshot restore for infected VM & 0.90 \\
& $a_{25}$: Clean DNS and FW rules & Remove temporary defense rules after validation & 0.15 \\
& $a_{26}$: Restore from backup & Recover critical files or configurations from safe backups & 0.80 \\
& $a_{27}$: Permanent hardening of policies & Persist firewall, auditd, and sudo settings to prevent recurrence & 0.20 \\
& $a_{28}$: Re-enable normal operation & Reconnect host to main VLAN after verification & 0.20 \\
\hline
\end{tabular}
\end{table*}

\begin{table*}[t]
\centering
\caption{PPO training parameters for the DeepStage DRL-based APT defense.}
\label{tab:ppo_params}
\small
\setlength{\tabcolsep}{4pt}
\begin{tabular}{p{4.0cm} p{4.5cm} p{8.5cm}}
\hline
\textbf{Parameter} & \textbf{Value / Range} & \textbf{Description} \\
\hline
Actor network & 3-layer MLP & Input: fused graph embedding ($g_t$) + stage belief ($p_t$); Output: probability over meta/sub-policy actions. \\
Critic network & 3-layer MLP & Estimates value $V(o_t)$ for baseline; shared first layer with actor. \\
Optimizer & Adam ($\beta_1{=}0.9$, $\beta_2{=}0.999$) & Adaptive optimization for actor and critic updates. \\
Learning rate ($\eta$) & $3\times10^{-4}$ (decay 0.99 per epoch) & Stable step size for PPO gradient updates. \\
Discount factor ($\gamma$) & $0.99$ & Long-term credit assignment for delayed APT mitigation rewards. \\
GAE parameter ($\lambda_{\text{GAE}}$) & $0.95$ & Balances bias and variance in advantage estimation. \\
Clipping threshold ($\epsilon$) & $0.2$ & Controls update size to maintain stable policy improvement. \\
Entropy coefficient ($c_{\text{ent}}$) & $0.01$--$0.05$ & Encourages exploration over meta/sub-policy choices. \\
Value loss coefficient ($c_v$) & $0.5$ & Balances critic loss contribution. \\
Batch size & 4096 transitions & Number of samples per PPO update. \\
Mini-batch size & 512 & Size for each gradient step in PPO optimization. \\
Epochs per update & 5 & PPO passes per batch; stabilizes on-policy learning. \\
Episode length & 50--100 time steps & Corresponds to one APT campaign (Caldera playbook). \\
Training episodes & 2,000 & Training and fine-tuning in the enterprise testbed. \\
Reward scaling & Normalize to $[-1,1]$ & Ensures numerical stability for PPO. \\
\hline
\end{tabular}
\end{table*}

\subsection{Controlled Testbed}
Experiments are conducted in a controlled enterprise testbed consisting of six Ubuntu 20.04 LTS virtual machines (VMs) representing enterprise hosts and one Kali Linux VM acting as the attacker. Each VM is provisioned with 4 vCPUs and 8\,GB RAM. A dedicated GPU workstation equipped with an NVIDIA RTX 3060 GPU and 32\,GB RAM is used for DRL training and computationally intensive tasks. Common enterprise services, including HTTP and SSH, are deployed across the host VMs, while a Zeek sensor~\cite{zeek_ids} is placed at the network edge to generate network-level alerts. Note that this testbed is designed as a proof-of-concept environment for evaluating the feasibility and effectiveness of the DeepStage. 

\subsection{Data collection and provenance}
Host telemetry is captured via \textit{auditd} and \textit{osquery}; kernel-level provenance is collected using CamFlow \cite{CamFlow}. Zeek produces network events, which are converted into Alert nodes and fused into the provenance graph (early fusion). Snapshots are constructed every $\Delta t = 300$ seconds to form $G_t$, and a GNN graph encoder produces embeddings $g_t\in\mathbb{R}^{128}$.

\subsection{Attack generation}
We use MITRE Caldera \cite{mitre_caldera} to generate realistic, ATT\&CK-aligned adversary behaviors for DRL training and evaluation. Caldera’s modular “ability” and “adversary” playbook model lets us script end-to-end multi-stage campaigns (reconnaissance$\rightarrow$initial compromise$\rightarrow$privilege escalation$\rightarrow$lateral movement$\rightarrow$C2$\rightarrow$exfiltration) and vary timing, techniques, and targets programmatically, which is ideal for producing large numbers of labeled episodes. In our pipeline, Caldera drives attacker actions against the enterprise VMs while host telemetry and network sensors collect the provenance. Each playbook execution is annotated by a controller that records ground-truth stage labels and timestamps, enabling per-snapshot labeling for stage-estimator training and reward shaping for the DRL agent. Specifically, we leverage 10 base Caldera playbooks \cite{mitre_caldera}, and generate randomized variants from each to increase variety and prevent overfitting.

\subsection{Meta-policy and sub-policies}
To enable realistic and fine-grained defense decisions within the emulated enterprise network, each meta-policy in the DeepStage framework is decomposed into multiple sub-policies that correspond to concrete, automatable defensive actions. These sub-policies are carefully selected to align with the system components available in the testbed, and can be executed safely through the management controller via REST automation on the target host. Table~\ref{tab:subpolicies} summarizes all practical sub-policies and their normalized operational costs $C(a)$, which are later incorporated into the DRL reward formulation. It is noted that the action $a_{0}$, representing the \emph{baseline monitoring mode}, is executed under normal system conditions when no attack is detected ($k$=0). 

\subsection{Reward computation}
The DRL agent receives a scalar reward, $R(b_t, a_t) = \alpha_k R_{\text{security}} - \beta_k \lambda C(a_t),$ at each time step $t$. In particular, $R_{\text{security}}(t) \in [0,1]$ quantifies the normalized \emph{security improvement} observed between time steps. Let $k_t$ and $k_{t+1}$ denote the ground-truth APT stages at time $t$ and $t{+}1$, respectively. Then, $R_{\text{security}}(t)$ is defined as
\begin{equation}
\label{eq:rsecurity_groundtruth}
R_{\text{security}}(t) =
\begin{cases}
1, & \text{if } k_{t+1} < k_t, \\
0.5, & \text{if } k_{t+1} = k_t, \\
0, & \text{if } k_{t+1} > k_t.
\end{cases}
\end{equation}
This piecewise formulation provides an intuitive and discrete reward signal: actions that successfully drive the adversary to an earlier or less critical stage yield the highest reward ($R_{\text{security}}=1$), while actions that merely stabilize the system provide partial credit ($R_{\text{security}}=0.5$). Conversely, if the attacker advances to a more severe stage, the reward drops to zero, signaling policy failure at time $t$. $C(a_t) \in [0,1]$ represents the normalized \emph{operational cost} of the selected action $a_t$, defined according to Table~\ref{tab:subpolicies}. The stage-specific coefficients $\alpha_k$ and $\beta_k$ are chosen to emphasize later-stage mitigation, where decisive defense actions are critical, while encouraging conservative low-cost monitoring during early stages. Typical settings are: $\alpha_k = [0.3,\, 0.5,\, 0.8,\, 1.0,\, 1.3,\, 1.5,\, 2.0]$ and $\beta_k = [0.0,\, 0.5,\, 0.7,\, 1.0,\, 1.3,\, 1.6,\, 2.0]$ for $k = 0, 1, \ldots, 6$, corresponding respectively to the Normal, Reconnaissance, Initial Compromise, Privilege Escalation, Lateral Movement, Command-and-Control, and Exfiltration stages. A typical cost scaling factor is set to $\lambda = 0.1$.

\subsection{DRL parameter settings} Table~\ref{tab:ppo_params} summarizes the hyperparameter settings and network configurations used to train the PPO-based reinforcement learning agent in the DeepStage framework. These parameters are selected to ensure stable policy convergence and effective learning under the partially observable and stage-specific APT defense \cite{PPO_Hyperparameters,Filali2022TNSE_DRL_RAN_Slicing}.
 
\subsection{Benchmarking Baselines}
To assess the performance of the proposed DeepStage framework, we benchmark it against the \textit{Risk-Aware} DRL-based APT Defense approach~\cite{AutomatedAPTDefense}. This baseline integrates reinforcement learning with MulVAL-derived attack-graph modeling, where each state encodes the risk levels associated with graph nodes and edges, and the action space comprises eleven predefined mitigation operations (e.g., patching, access restriction, or service isolation). The reward function is defined as the net reduction in aggregate network risk minus the operational cost of the selected mitigation, thereby optimizing risk minimization under constrained defensive budgets. This formulation represents a risk-centric defense paradigm that leverages symbolic attack-graph reasoning rather than data-driven behavioral modeling.

Furthermore, we include a stage-unaware variant of our framework, denoted \textit{DeepStage-Unaware}, to quantify the contribution of stage conditioning to overall performance. In this variant, the stage-specific weighting factors in the reward function are neutralized by setting $\alpha_k = \beta_k = 1.0$, resulting in a uniform reward across all APT phases. Consequently, DeepStage-Unaware learns a defense policy without explicit awareness of the attacker’s progression along the kill chain, allowing direct evaluation of the benefit provided by the proposed stage-aware reinforcement structure.

\subsection{Result Analysis}

\begin{table*}[t]
\centering
\caption{Per-stage F1-score and overall defense effectiveness comparison.}
\label{tab:defense_results}
\small
\begin{tabular}{lcccccccc}
\hline
Method & Recon & InitAcc & PrivEsc & LatMov & C2 & Exfil & Avg-F1 & Mitigation Success (\%) \\
\hline
Risk-Aware DRL       & 0.75 & 0.73 & 0.70 & 0.71 & 0.76 & 0.72 & 0.728 & 68.5 \\
DeepStage-Unaware    & 0.82 & 0.79 & 0.76 & 0.78 & 0.83 & 0.80 & 0.797 & 76.7 \\
DeepStage            & \textbf{0.91} & \textbf{0.88} & \textbf{0.85} & \textbf{0.87} & \textbf{0.92} & \textbf{0.89} & \textbf{0.887} & \textbf{84.7} \\
\hline
\end{tabular}
\end{table*}

\paragraph{Attack Defense Performance}
Table~\ref{tab:defense_results} compares the per-stage defense effectiveness of all methods across the APT lifecycle, together with their mitigation success rates. DeepStage achieves the best results in all stages, with an average F1-score of 0.887, outperforming DeepStage-Unaware and the Risk-Aware DRL baseline by approximately 11.3\% and 21.8\%, respectively. The gains are consistent across both early and late attack phases. In particular, DeepStage achieves F1-scores of 0.87 and 0.92 for Lateral Movement and Command-and-Control, where attack behaviors often span multiple hosts and produce weak but correlated indicators. This shows that combining provenance graph embeddings with probabilistic stage beliefs improves the policy's ability to capture attack progression and adapt responses accordingly. Compared with DeepStage-Unaware, DeepStage benefits from stage-conditioned reward weights $(\alpha_k,\beta_k)$, which encourage low-cost monitoring in early stages and stronger containment or remediation in high-risk stages. This enables more timely and stage-appropriate defense decisions.

DeepStage also substantially improves attack mitigation effectiveness. Its mitigation success rate reaches 84.7\%, compared with 76.7\% for DeepStage-Unaware and 68.5\% for Risk-Aware DRL. Here, mitigation success rate is defined as the percentage of attack episodes in which the defense agent prevents the adversary from reaching critical stages, such as Command-and-Control or Exfiltration. The 16.2\% improvement over the Risk-Aware DRL baseline demonstrates that the proposed hierarchical PPO agent can select more timely and effective defensive actions than policies based on static attack-graph risk abstractions. Overall, these results confirm that DeepStage improves both per-stage defense accuracy and end-to-end APT mitigation effectiveness.

\begin{figure}[t]
    \centering
    \includegraphics[width=0.5\textwidth]{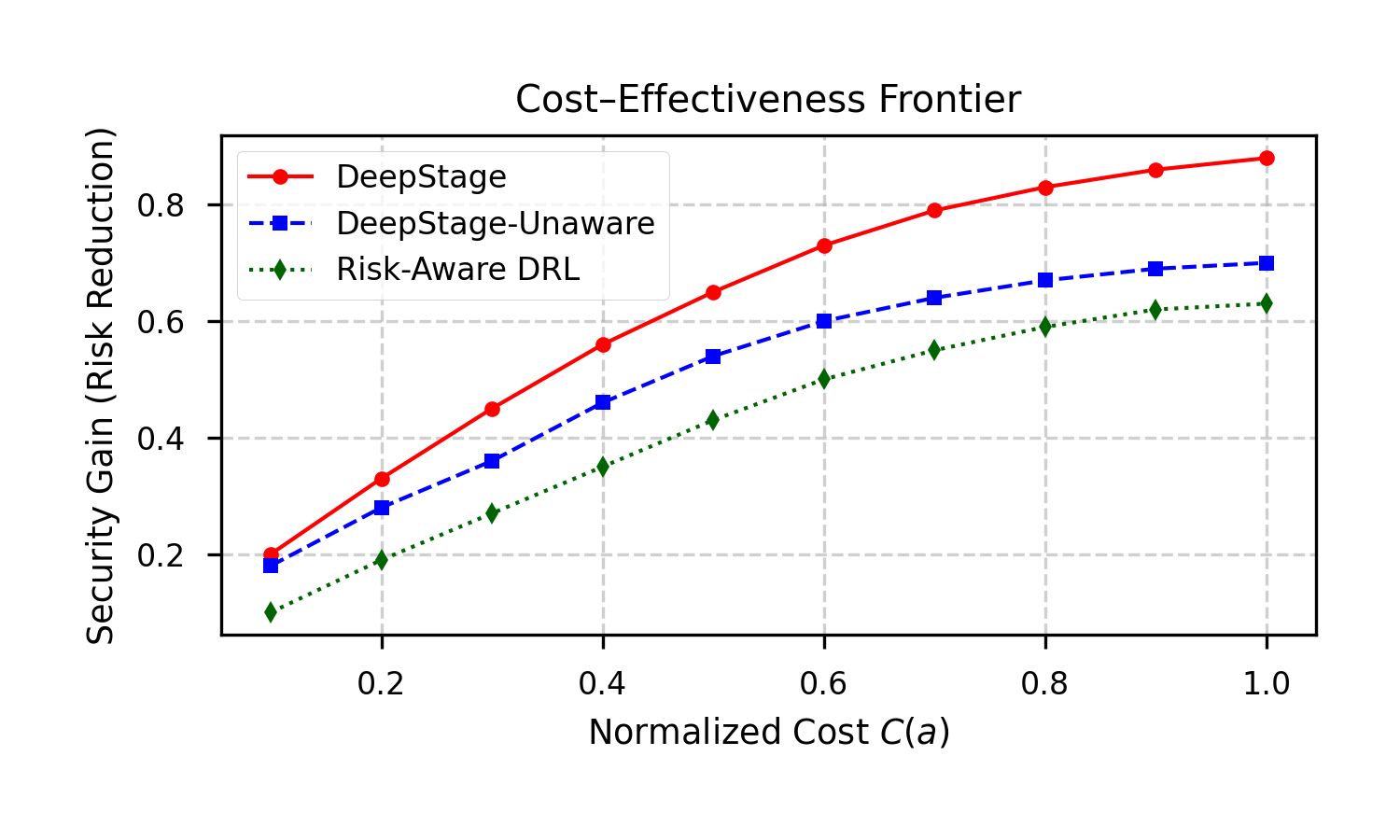}
    \caption{Cost–effectiveness frontiers illustrating normalized security gain versus cumulative action cost.}
    \label{fig:cost_frontier}
\end{figure}

\paragraph{Cost--Effectiveness Analysis}
The cost--effectiveness frontier in Figure~\ref{fig:cost_frontier} demonstrates that DeepStage achieves consistently higher security gain for comparable operational cost levels. At the mid-range cost ratio ($C(a)/C_{\max}=0.5$), DeepStage attains a normalized security gain of 0.65, outperforming DeepStage-Unaware (0.54) and the Risk-Aware DRL baseline (0.43). This represents an average improvement of approximately 23\% and 51\%, respectively. Even under full budget allocation, DeepStage maintains a 15--25\% margin in cumulative reward, indicating that the hierarchical PPO agent effectively balances mitigation cost and defensive value. The stage-aware reward shaping $(\alpha_k, \beta_k)$ enables strategic resource allocation---favoring lightweight monitoring during early APT stages and escalating to containment or remediation only when high-confidence late-stage indicators emerge. In contrast, the stage-unaware variant triggers redundant actions in low-risk stages, while the Risk-Aware baseline remains overly conservative due to its coarse attack-graph abstraction.

\begin{figure}
    \centering
    \includegraphics[width=0.5\textwidth]{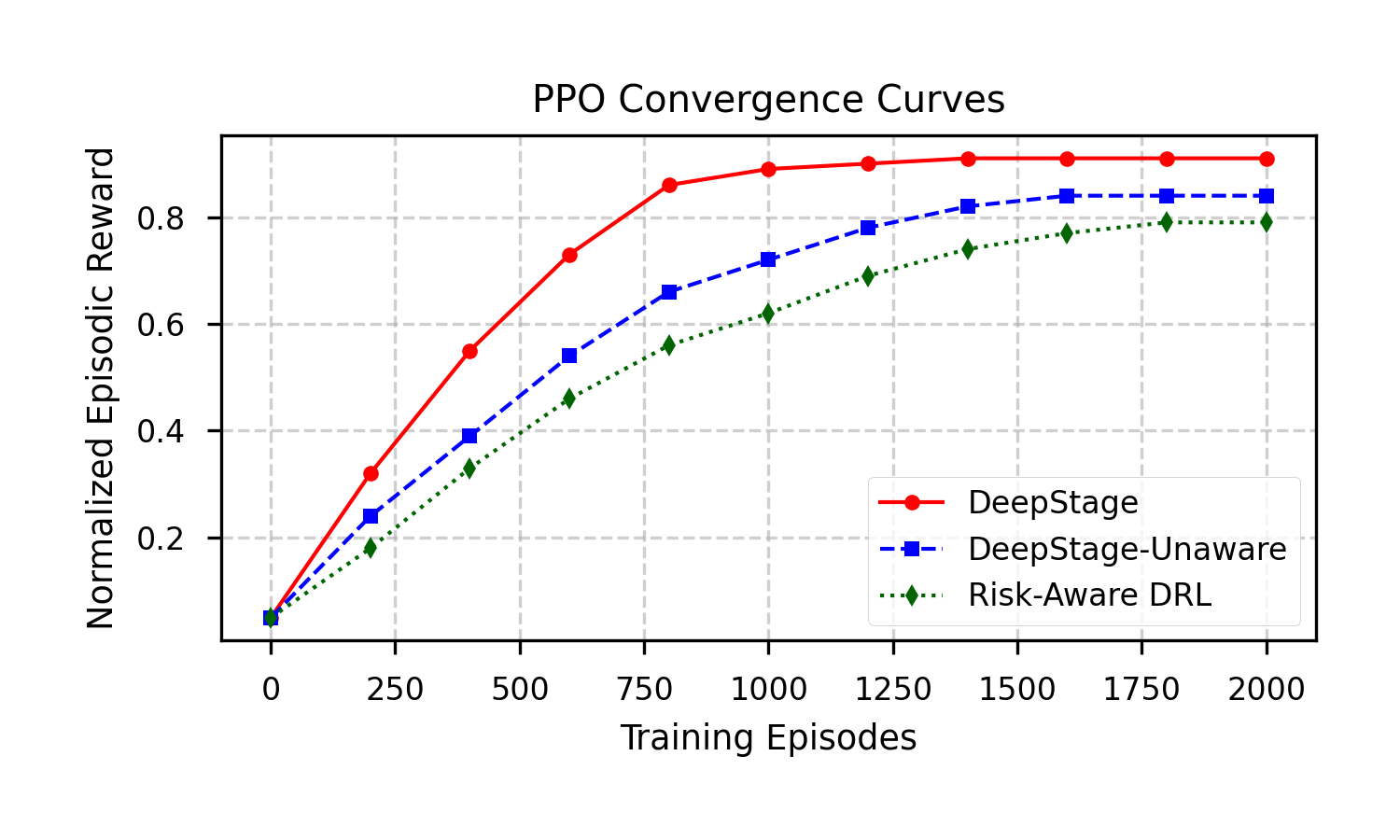}
    \caption{Training convergence of hierarchical PPO across methods.}
    \label{fig:ppo_convergence}
\end{figure}

\paragraph{Learning Stability and Convergence}
Figure~\ref{fig:ppo_convergence} shows the PPO learning curves for the three evaluated methods. DeepStage converges approximately 1.7$\times$ faster than both baselines, reaching a stable policy after about 900 episodes, while DeepStage-Unaware and the Risk-Aware DRL baseline require around 1600 and 1800 episodes, respectively. At convergence, DeepStage achieves a normalized episodic reward of 0.91, compared with 0.84 for DeepStage-Unaware and 0.79 for the Risk-Aware model, yielding an average improvement of roughly 12\%. This acceleration and stability arise from DeepStage’s hierarchical PPO design, which decomposes the defense policy into four structured meta-policies, reducing the effective action-space complexity and enabling faster, more stable updates. The stage-conditioned reward shaping further enhances learning efficiency by providing phase-specific reinforcement aligned with the attacker’s progression. By contrast, the Risk-Aware baseline exhibits a flatter and slower convergence curve due to its static, low-dimensional attack-graph state representation, which limits gradient feedback and delays policy stabilization.

\begin{figure}
    \centering
    \includegraphics[width=0.5\textwidth]{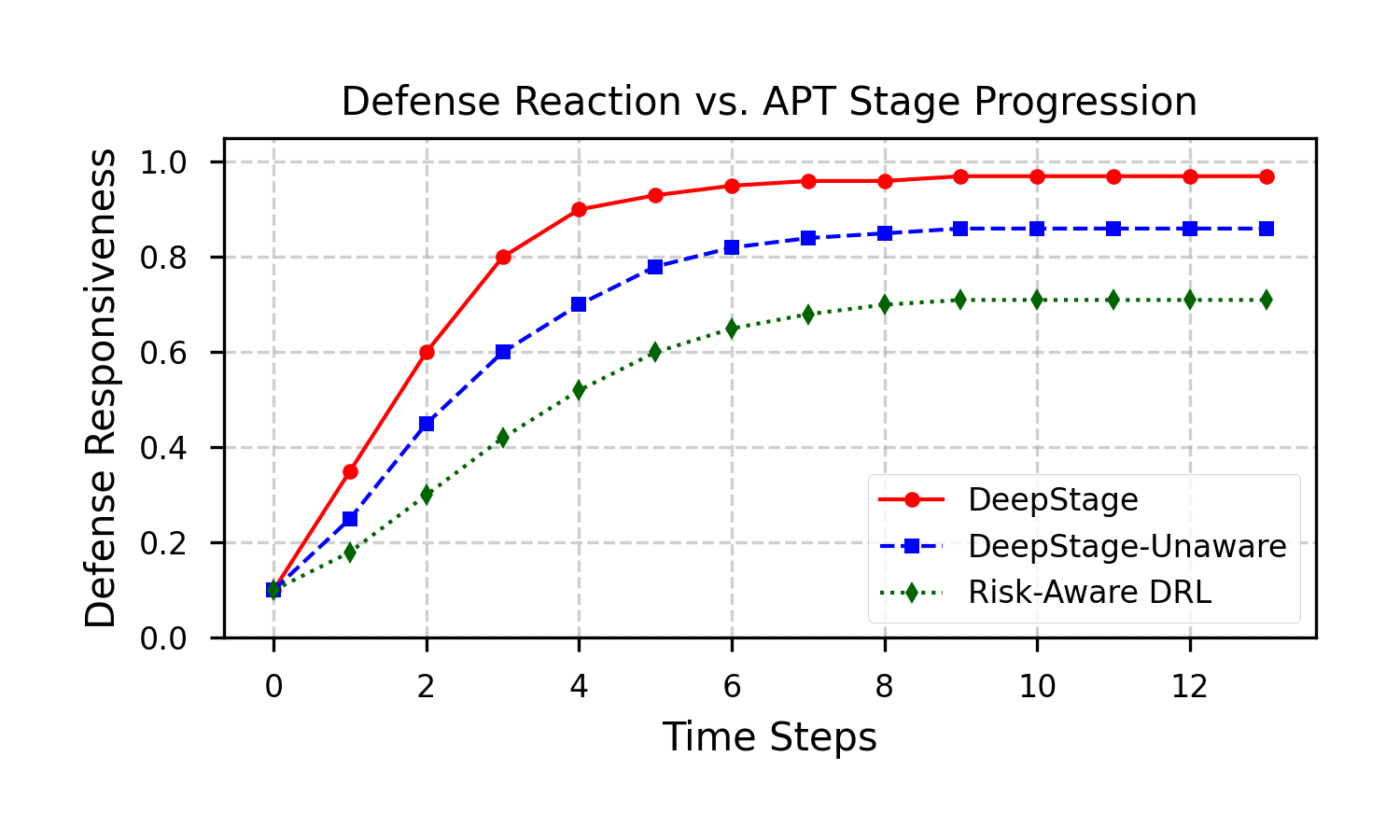}
    \caption{Defense responsiveness over APT stage transitions.}
    \label{fig:stage_timeline}
\end{figure}

\paragraph{Stage-Transition Responsiveness}
Figure~\ref{fig:stage_timeline} shows the defense responsiveness of the three methods over successive time steps, where each step corresponds to a 300-second observation window. 
DeepStage exhibits the most agile and timely adaptation, reaching a responsiveness level of 0.90 by the fourth time step (approximately 20 minutes of emulated time) and stabilizing near 0.97 thereafter. 
In contrast, DeepStage-Unaware attains a maximum responsiveness of 0.86, while the Risk-Aware DRL baseline saturates around 0.71. 
This demonstrates that the stage-aware reinforcement mechanism in DeepStage enables rapid escalation of mitigation actions as the attack progresses through successive APT stages. 
The integration of the LSTM-based stage estimator with the hierarchical PPO agent ensures that higher-confidence stage transitions trigger more assertive defensive policies, allowing the system to contain or remediate intrusions early. 
By comparison, the stage-unaware variant exhibits delayed responses during early- and mid-stage transitions, and the Risk-Aware baseline responds conservatively due to its static attack-graph abstraction and limited temporal sensitivity.

\section{Conclusion}\label{Conclusion}

This paper introduced \textit{DeepStage}, a unified deep reinforcement learning framework for adaptive and stage-aware defense against Advanced Persistent Threat (APT) attacks. DeepStage fuses multi-modal host and network provenance data into graph-based representations that capture structural and behavioral dependencies across enterprise systems. These fused graphs are encoded into low-dimensional embeddings and processed by an LSTM-based stage estimator to infer probabilistic beliefs of the attacker’s current phase along the kill chain. The resulting stage probabilities, combined with graph embeddings, define the state representation for a hierarchical PPO agent that organizes the defense policy into four interpretable layers: monitoring, access control, containment, and remediation. 

Through this design, DeepStage enables coordinated and context-aware defense actions that adapt to the evolving stages of sophisticated multi-stage attacks. Experimental results demonstrate the effectiveness of the proposed framework in accurately estimating attack stages and improving the overall effectiveness of autonomous mitigation strategies. Overall, DeepStage provides a principled and extensible foundation for autonomous APT defense by integrating situational awareness, temporal reasoning, and resource-aware decision-making in enterprise security environments.

\section{Future Research Directions}\label{Futurework}
As future work, we plan to extend DeepStage to improve the transparency and operational usability of autonomous cyber defense. First, we aim to incorporate explainable artificial intelligence (XAI) techniques to provide interpretable insights into attack-stage estimation and defense policy decisions. By identifying critical provenance subgraphs, influential system events, and key features driving mitigation actions, XAI can improve analyst trust and support evidence-driven verification of automated responses. Explanation signals may also guide reinforcement learning optimization, aligning defense policies more closely with interpretable attack evidence. Second, we plan to integrate large language model (LLM)-based security operations center (SOC) assistants that translate DeepStage outputs into human-readable incident reports, attack timelines, and mitigation rationales. These capabilities can improve human--AI collaboration by helping analysts quickly understand alerts and defense recommendations. Together, these directions aim to bridge autonomous cyber defense and practical security operations, enabling more trustworthy protection against multi-stage APT campaigns.

\section*{Acknowledgment}
This work has been performed in the framework of the SUSTAINET-Advance project, funded by the German BMFTR (ID:16KIS2280).

\bibliographystyle{ieeetr}
\bibliography{References.bib}

\end{document}